\documentclass[sigconf]{acmart}

\def\BibTeX{{\rm B\kern-.05em{\sc i\kern-.025em b}\kern-.08emT\kern-.1667em\lower.7ex\hbox{E}\kern-.125emX}}
    
\setcopyright{rightsretained}

\usepackage{booktabs} 
\usepackage{url}
\usepackage{algorithm}
\usepackage[noend]{algpseudocode}
\makeatletter
\def\BState{\State\hskip-\ALG@thistlm}
\makeatother

\usepackage{etoolbox,siunitx}
\robustify\bfseries
\usepackage{algpseudocode}
\usepackage{arydshln}

\usepackage{color}
\usepackage{hyperref}
\hypersetup{
    unicode=false,          
    pdftoolbar=true,        
    pdfmenubar=true,        
    pdffitwindow=false,     
    pdfstartview={FitH},    
    pdfauthor={Farhan Khawar},     
    pdfsubject={Subject},   
    pdfcreator={Farhan Khawar},   
    pdfproducer={Producer}, 
    pdfnewwindow=true,      
    colorlinks=true,       
    linkcolor=red,          
    citecolor=blue,        
    filecolor=magenta,      
    urlcolor=cyan           
}

\begin{document}

\title{Cleaned Similarity for Better Memory-Based Recommenders}

\author{Farhan Khawar}
\email{fkhawar@cse.ust.hk}
\affiliation{%
  \institution{Department of Computer Science \& Engineering\\ The Hong Kong University of Science and Technology}
  \city{Hong Kong}
}

\author{Nevin L. Zhang}
\email{lzhang@cse.ust.hk}
\affiliation{%
  \institution{Department of Computer Science \& Engineering\\ The Hong Kong University of Science and Technology}
  \city{Hong Kong}}

%
\renewcommand{\shortauthors}{F. Khawar and N. L. Zhang}

\begin{abstract}

Memory-based collaborative filtering methods like user or item k-nearest neighbors (kNN) are a simple yet effective solution to the recommendation problem. The backbone of these methods is the estimation of the empirical similarity between users/items. In this paper, we analyze the spectral properties of the Pearson and the cosine similarity estimators, and we use tools from random matrix theory to argue that they suffer from noise and eigenvalues spreading. We argue that, unlike the Pearson correlation, the cosine similarity naturally possesses the desirable property of eigenvalue shrinkage for large eigenvalues. However, due to its zero-mean assumption, it overestimates the largest eigenvalues. We quantify this overestimation and present a simple re-scaling and noise cleaning scheme. This results in better performance of the memory-based methods compared to their vanilla counterparts.
\end{abstract}

%
%


\maketitle
\section{Introduction}

Collaborative Filtering (CF) methods are one type of recommendation techniques that use the \emph{past} interactions of other users to filter items for a single user. Broadly speaking, CF methods are generally characterized into memory-based and model-based methods. Memory-based methods are known for their simplicity and competitive performance \cite{volkovs2015effective}. Recently, they have been successfully used for session-based recommendations\cite{Jannach:2017:RNN:3109859.3109872} and they are still used as a part of the recommendation solution in industry\cite{slack}.

Memory-based methods like user-kNN and item-kNN extract user (or item) similarities which are used to form user (or item) neighborhoods by taking the $k$-nearest neighbors. These neighborhoods are then used to filter items for a user. 

Calculating the similarity effectively is of great importance in these methods. One of the most commonly used similarity metrics is cosine similarity. Formally, the cosine similarity between two users $x$ and $y$ can be defined as: 

\begin{equation}
\label{eq:cosine}
\sigma = \frac{\sum_{i=1}^{n}x_iy_i}{ \sqrt{ \sum_{i=1}^{n} x_i^2}  \sqrt{ \sum_{i=1}^{n} y_i^2} },
\end{equation} 
where, $n$ is the total number of samples (items in this case) and $x_i$ and $y_i$ represent the preferences of user $x$ and user $y$ on the $i$-th item respectively. The similarity between two items is defined in a similar manner. If the data is centered then the cosine similarity is equivalent to the empirical correlation which is calculated by:
\begin{equation}
\label{eq:pearson}
\sigma = \frac{\sum_{i=1}^{n}(x_i-\bar{x})(y_i-\bar{y})}{ \sqrt{ \sum_{i=1}^{n} (x_i-\bar{x})^2}  \sqrt{ \sum_{i=1}^{n} (y_i-\bar{y})^2} },
\end{equation} 
where, $\bar{x}$ is the sample mean i.e., $\frac{1}{n}\sum_{i=1}^{n}x_i$, and analogously for $\bar{y}$.

The empirical correlation, and hence the cosine similarity, is a good estimation of the true correlation when the number of samples is large. However, in practice the number of users is of the same order as the number of items and the ratio of the number of users to the number of items is not very small compared to 1. In this case, the empirical correlations are dominated by noise and care should be taken while using them as similarities.

The correlations between users (or items) can be viewed as an empirical correlation matrix where each entry denotes the empirical correlation of the entities represented by its index e.g., the entry at the index $(1,5)$ of the user empirical correlation matrix would be the correlation between user 1 and user 5.
Results from random matrix theory (RMT) can then be used to understand the structure of the eigenvalues and eigenvectors of this empirical correlation matrix. The main contributions of this paper are as follows:
\begin{itemize}
\item We analyze the structure and spectral properties of the Pearson and cosine similarity. 
\item We argue that Cosine similarity possesses the desirable property of eigenvalue shrinkage.
\item We quantify the overestimation of the largest eigenvalue in cosine similarity.
\item We show that the theoretical results regarding the distribution of eigenvalues of random matrices can be used to clean the noise from the empirical user/item correlation matrix.
\end{itemize}
  
\section{Preliminaries of RMT}\label{PRMT}
RMT theorems attempt to make statements about the spectral properties of large random correlation matrices \footnote{RMT theorems are also applicable to other general matrices.}. They are applied in the case when an $n  \times m$  random matrix $\mathbf{X}$  with independent and identically distributed (i.i.d.) random entries of zero-mean  is such that $m,n \rightarrow \infty$ and the ratio $m/n \rightarrow q \in (0,1]$.

Interestingly, the eigenvalue distribution of the empirical correlation matrix of $\mathbf{X}$ is known exactly under these conditions and given by the Mar$\check{c}$enko Pastur law (MP-law):
\begin{equation}
\label{RMT}
\rho_{\mathbf{X}}(\lambda) = \frac{1}{2 \pi q \lambda} \sqrt{(\lambda_{max}-\lambda)(\lambda-\lambda_{min})},
\end{equation}
where the eigenvalue $\lambda \in [\lambda_{max}, \lambda_{min}] $ and $\lambda_{max} = (1+\sqrt{q})^2$ and $\lambda_{min} = (1-\sqrt{q})^2$.

\begin{figure}
		\centering
		\includegraphics[scale=0.2]{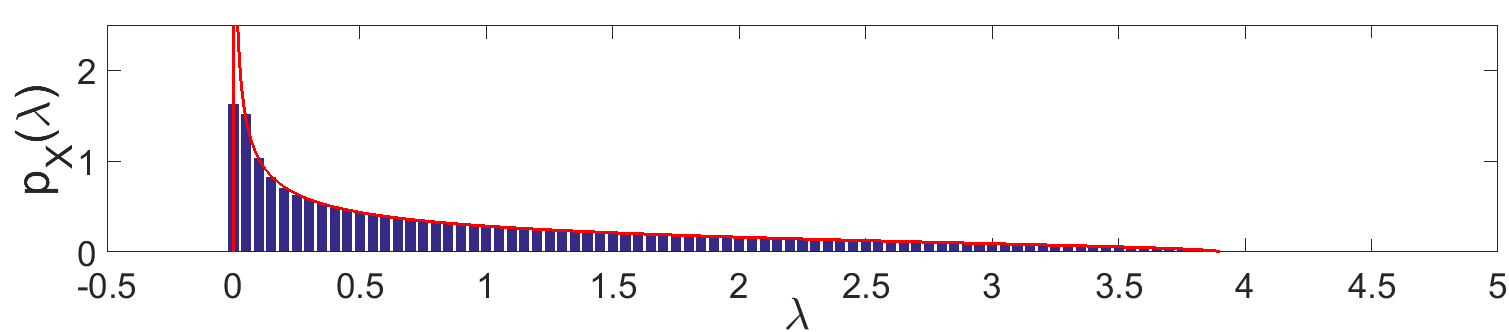}
		\caption{The solid line shows the plot of the MP-law density from Equation \ref{RMT}. The histogram obtained from eigenvalues of a random matrix follows the MP-law distribution. }
		\label{fig:RMTrand}
	\end{figure}

This result implies that there should be no eigenvalues outside the interval $ [\lambda_{max}, \lambda_{min}]$ for a random noise correlation matrix. A plot of the density of Equation \ref{RMT} is shown in Figure \ref{fig:RMTrand} along with the eigenvalue distribution of a random item correlation matrix formed by randomly permuting the entries of each column of a user-item feedback matrix. As we can see the histogram follows the theoretical MP-law distribution quite accurately.

\section{Cleaning the correlation matrix}\label{cleaning}
	Using the result where a pure noise correlation matrix has an eigenvalue distribution similar to MP-law in the limiting case, we can clean the user (or item) correlation matrix by comparing its empirical eigenvalue distribution with that of the MP-law. If the bulk of the eigenvalues are within the range $[\lambda_{max}, \lambda_{min}]$ and their distribution resembles the MP-law then it is most probably due to noise and can be ignored.
	
	A simple strategy is to remove all eigenvalues between  RMT ``noise bulk'' range i.e., $[\lambda_{min},\lambda_{max}]$ by setting them to 0, and retaining the rest of the eigenvalues. However, in practice the eigenvalue distribution in the noise bulk range does not follow the MP-law exactly. Therefore, a cutoff point near $\lambda_{max}$ is used instead of $\lambda_{max}$. This cutoff point $\lambda_{cut}$ is usually searched within a range near $\lambda_{max}$. This strategy is known as eigenvalue clipping \cite{bouchaud2009financial}.
	
\subsection{Eigenvalue spreading}
The empirical correlation estimator of Equation \ref{eq:pearson}, also known as the Pearson or the sample correlation matrix is a common estimator of the true user or item correlation matrix. When we have a much larger number of datacases compared to the number of features i.e., $q \rightarrow 0$ then this estimator approaches the true correlation matrix. However, when the number of datacases and the number of features are of the same order i.e., $q = O(1)$, the MP-law states that the empirical correlation estimate becomes a noisy estimate of the true correlation matrix. This is because if the true correlation matrix is an identity matrix (pure noise) then the distribution of the eigenvalues of the empirical correlation is not a single spike at 1, but rather it is spread out as shown in Figure \ref{fig:RMTrand}. This spreading out is dependent on $q$ itself and given by the MP-law stated in Equation \ref{RMT}. The spectrum gets more spread out (noisier) as q increases. This tells us that when we have a data sample in the regime $q =O(1)$ then the small eigenvalues are \emph{smaller} and the large eigenvalues are \emph{larger} compared to the corresponding eigenvalues of the true correlation matrix. Therefore, the cleaning strategy should take this into account and shrink the estimated eigenvalues appropriately.
\subsection{Zero-mean assumption}
The Pearson estimator is more general as it assumes that the data is not-zero mean, which is often the case in practice. However, the data in collaborative filtering are large and sparse, and applying the Pearson correlation estimator on this data would imply making this large user-item matrix $\mathbf{X}$ dense (by removing the mean from each entry of the matrix). This is problematic from both the memory and computational points of view. 

The MP-law was stated for the zero-mean data. The Pearson estimator standardizes the data to make it zero-mean, therefore we can use the MP-law results. In this subsection, we show that we can use the findings from MP-law for the case when the data is not zero-mean. This is because any matrix $\mathbf{X}$ can be written as:
\begin{equation}
\label{eq:mean}
\tilde{\mathbf{X}}   = \mathbf{X} - \mathbf{M},
 \end{equation}
 where, $\tilde{\mathbf{X}}$ is the demeaned  version of $\mathbf{X}$ and $\mathbf{M} = \mathbf{1}_n \times \mathbf{m}$ is an $n \times m$ matrix, where each row is equal to the vector $\mathbf{m}$. Additionally, $\mathbf{m}$ is a $1 \times m$ row vector that contains the column mean of the corresponding columns of $\mathbf{X}$ and $\mathbf{1}_n$ is a $1 \times n$ vector of all 1's. 
Then we can rewrite the Pearson correlation estimation as:
\begin{equation}
\label{eq:pearsonnew}
\mathbf{E}_p = \frac{1}{n}\tilde{\mathbf{X}}^T\tilde{\mathbf{X}} = \frac{1}{n}(\mathbf{X}^T\mathbf{X} - \mathbf{M}^T\mathbf{M}),
 \end{equation}

where, w.l.o.g., for simplicity of notation, we assume that data has unit variance. It is trivial to see that $\mathbf{M}^T\mathbf{M}$ is of rank 1 and has one eigenvalue $\xi $, which is a positive number.
We know from the subadditivity property of rank that:

\begin{align}
\label{eq:rankUB} 
 rank(\mathbf{X}^T\mathbf{X})&=rank(\tilde{\mathbf{X}}^T\tilde{\mathbf{X}} + \mathbf{M}^T\mathbf{M} )\\ &\leq rank(\tilde{\mathbf{X}}^T\tilde{\mathbf{X}}) + rank(\mathbf{M}^T\mathbf{M}),\\ 
  & \leq   N +1,  
\end{align}
 
 where, $rank(\tilde{\mathbf{X}}^T\tilde{\mathbf{X}})=N$ and it can also be shown \cite{944751} that since $rank( \mathbf{M}^T\mathbf{M})=1$ then:
 \begin{equation}
\label{eq:rankLB} 
 rank(\mathbf{X}^T\mathbf{X})=rank(\tilde{\mathbf{X}}^T\tilde{\mathbf{X}} + \mathbf{M}^T\mathbf{M} ) \geq N -1, 
 \end{equation}
therefore, the rank of the correlation matrix ($\frac{1}{n}\mathbf{X}^T\mathbf{X}$) of data will change by at most 1, if at all, compared with the rank of the correlation matrix of the demeaned data. As we will see next, the eigenvalue $\xi $ is positive and large, so it will \emph{only} affect the top eigenvalues of the correlation matrix of the original data. 
  
\begin{figure}[]
	
		\includegraphics[scale=.20]{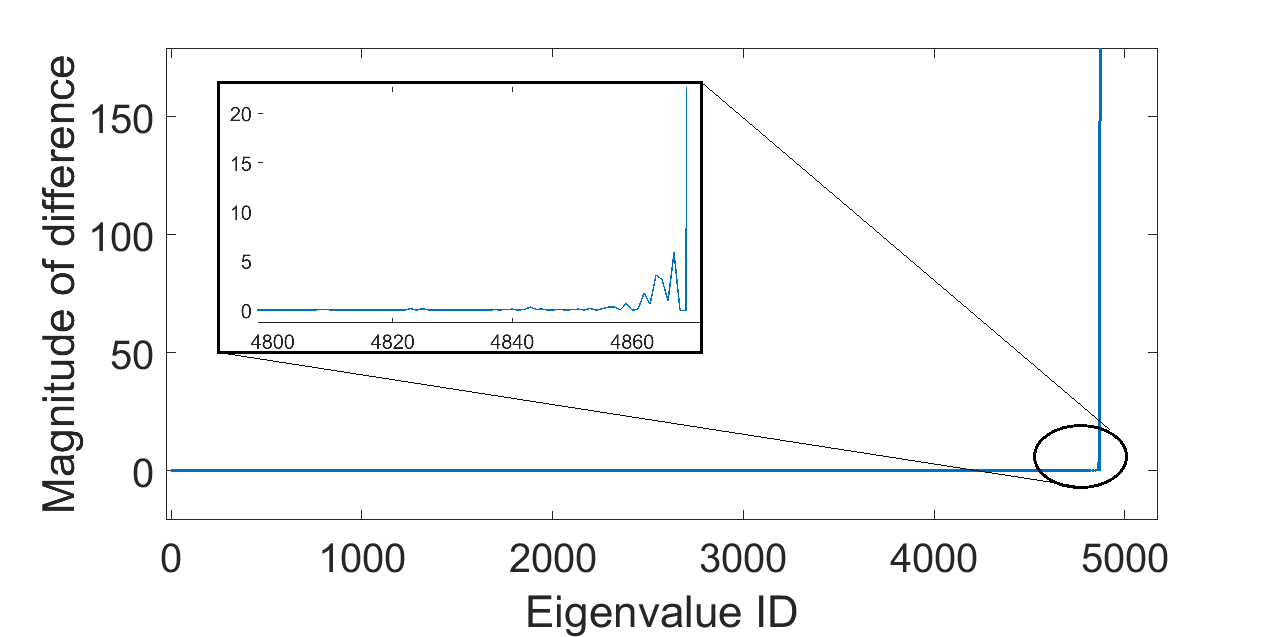}

		\caption{The magnitude of the difference in the corresponding eigenvalues of the original data correlation matrix and de-meaned data correlation matrix is shown on the y-axis, against the ID of the eigenvalue on the x-axis.}
		\label{fig:diff1}
	\end{figure}

 In Figure \ref{fig:diff1} we plot the \emph{difference} in the eigenvalue magnitudes of the user correlation matrices of the original data and the demeaned data for the Movielens1M dataset, where the eigenvalues of both matrices are sorted in the ascending order of magnitude. We can see a huge positive spike at the largest eigenvalue, signifying that the largest eigenvalue of the original data correlation matrix is overestimated, and a couple of relatively negligible spikes. From the discussion in the previous subsection, the largest eigenvalue of the demeaned data correlation matrix is already overestimated and the effect of not removing the mean exaggerates it further. Therefore, the effect of not removing the mean from the data is that the largest eigenvalue of the correlation matrix is overestimated.
 
In the context of recommender systems, where the data are sparse and large, this means that we can operate on the sparse data matrices by correcting for this overestimation. Moreover, since not demeaning the data effectively just changes the top eigenvalue, we can still use the eigenvalue clipping strategy and other insights based on the MP-law. 
\subsection{Quantifying the overestimation}
Interestingly this overestimation can be quantified by the eigenvalue of $\frac{1}{n}\mathbf{M}^T\mathbf{M}$. The sum of the difference shown in Figure \ref{fig:diff1} is exactly equal to $\xi $. This is trivially true since the trace of the data correlation matrix is to be preserved.

We do not need to do the eigenvalue decomposition of $\frac{1}{n}\mathbf{M}^T\mathbf{M}$  to get $\xi $. This is because, firstly, the eigenvalue of a rank 1 matrix is equal to its trace by the following argument; $\frac{1}{n}\mathbf{M}^T\mathbf{M}=uv^T$ is an $m \times m$ rank 1 matrix, where $u,v$ are $m \times 1$ vectors. Since $m \geq 1$ the matrix is singular and has 0 as its eigenvalue. We know if $\mu$ is the eigenvector associated with $\xi $ then:
\begin{align}
(uv^T) \mu & = \xi  \mu,\\
u(v^T \mu) / \xi & = \mu,
\end{align}
since $(v^T \mu) / \xi$ is a scalar, $u$ is also an eigenvector associated with $\xi$. Then, it follows that $u(v^T u) = \xi  u$, and as $u \neq 0$ we have $\xi = (v^T u) = \sum_{i=1}^{m}v_iu_i = Tr(\frac{1}{n}\mathbf{M}^T\mathbf{M}) $. Secondly, the trace of $\frac{1}{n}\mathbf{M}^T\mathbf{M}$ is non-zero by the construction of the matrix $\mathbf{M}$.

The matrix$\frac{1}{n}\mathbf{M}^T\mathbf{M}$ is dense and when $m$ is large calculating this matrix gets unfeasible. However, we notice that we are only interested in the diagonal of the above matrix and not the complete matrix. Therefore, the above trace can efficiently be calculated by:
\begin{equation}
Tr(\frac{1}{n} \mathbf{M}^T\mathbf{M})  =
\sum_{i=1}^{m} n \tilde{m}_i^2, 
\label{eqn:overestimate}
\end{equation}
where, $\tilde{m}_i = m_i/ \sqrt{n}$ and $m_i$ is the $i-th$ element of $\mathbf{m}$. Equation \ref{eqn:overestimate} gives us an efficient way to quantify the overestimation in the top eigenvalue of $\mathbf{X}^T\mathbf{X}$ \footnote{The discussion so far generalizes to the case when columns of $\mathbf{X}$ are not a unit variance by dividing each column of $\mathbf{X}$ and $\mathbf{M}$ by the standard deviation of the corresponding column of $\mathbf{X}$.}.

\subsection{Eigenvalue shrinkage}
Before we outline our cleaning procedure we briefly talk about cosine similarity. Cosine similarity assumes that the data is zero mean, however, this is not true in general. Moreover, based on our previous discussion, it does not make the correction for this by scaling the largest eigenvalue. 

\begin{figure}[]
	
		\includegraphics[scale=.20]{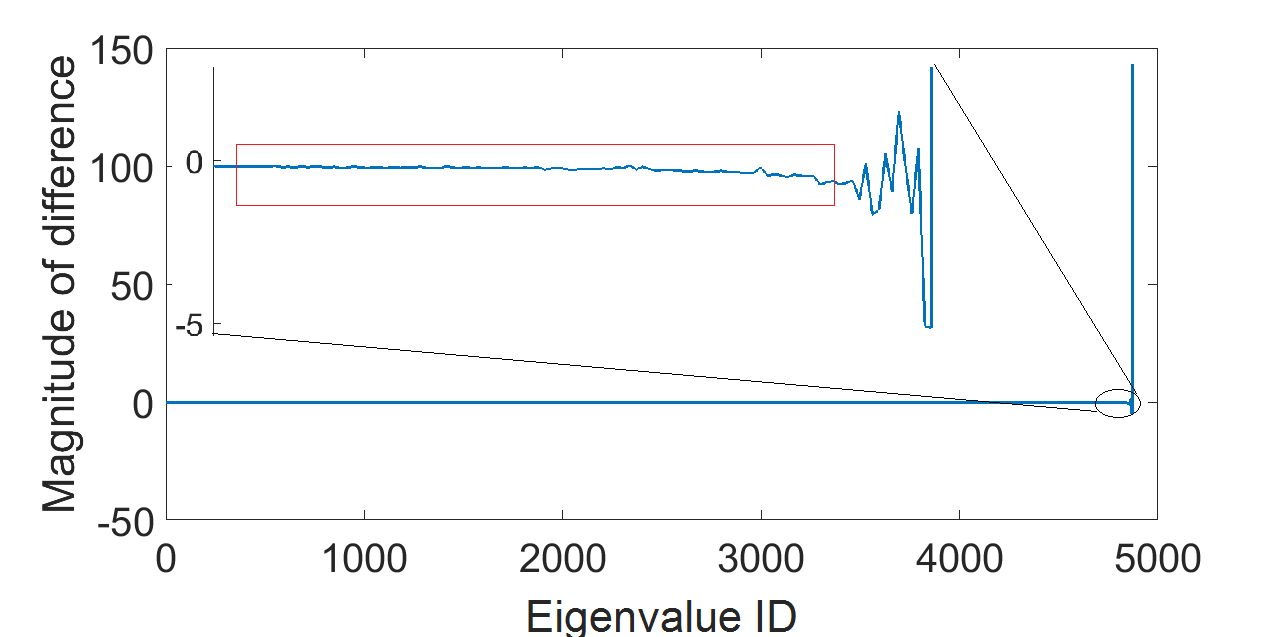}

		\caption{The the magnitude of the difference in the corresponding eigenvalues of the Pearson correlation matrix and Cosine correlation matrix is shown. The negative slope, highlighted by the red box, signifies the shrinkage property of cosine similarity.}
		\label{fig:diffc1}
	\end{figure}

However, when we plot the difference in the eigenvalues of the cosine similarity and the Pearson correlation, we find some interesting results. As seen in Figure \ref{fig:diffc1}, we have a large spike at the top eigenvalue as before which is expected since cosine similarity does not remove the mean. This is followed by some oscillations, but these oscillations are negative too. This can be due to the difference in variance. Finally, and more importantly, unlike before, the difference between the magnitude of eigenvalues of cosine similarity and Pearson correlation for all the other top eigenvalues is not very close to 0. In fact, we can see a gradual upward slope in the zoomed-in plot in Figure \ref{fig:diffc1} which was not visible before. 

This negative slope signifies that the top eigenvalues of cosine similarity (except the maximum eigenvalue) are shrunk compared to the eigenvalues of the Pearson correlation. Therefore, the cosine similarity implicitly does eigenvalue shrinkage.

The reason for this shrinkage is that the column variances of the data calculated in the Pearson correlation and cosine similarity are not the same. This can been seen from the denominators of Equation \ref{eq:cosine} and Equation \ref{eq:pearson}. When this is the case we cannot write a simple expression like Equation \ref{eq:pearsonnew} since both matrices on the right-hand side will have different column variances(the $\mathbf{M}^T \mathbf{M}$ matrix comes from the Pearson correlation). Consequently, the simple analysis that followed will not hold, hence the effect of not removing the mean will be more complex and in this case in the form of shrinkage of the top eigenvalues except the maximum eigenvalue.

\subsection{Cleaning  algorithm}
Below we outline a linear time and memory efficient similarity matrix cleaning strategy that explicitly shrinks the top eigenvalue, inherits the shrinkage property of cosine similarity for other eigenvalues\footnote{This shrinkage(both explicit and inherent) is not present in vanilla SVD/PCA.} and removes noise by clipping the smaller eigenvalues. 

\begin{algorithm}
\caption{Clean-KNN($\mathbf{X}$,$F$)}\label{algo}
\footnotesize
\begin{description}
\item[Inputs:] Sparse user-item matrix $\mathbf{X}$,, number of top eigenvalues $F$. \\
\end{description}
\begin{algorithmic}[1]
\Procedure{Learn Item-Item Similarity}{}
\BState \emph{One-pass over non-zero entries:}
\State  Calculate column mean vector $\mathbf{m}$;
\State Calculate column sum vector $\mathbf{\sigma}$;
\BState \emph{One-pass over the non-zero entries} $x_{ij}$ \emph{of} $\mathbf{X}$:
\State $\mathbf{X'}=[x_{ij}/ \sigma_j]_{ij}$, divide each $x_{ij}$ by its column sum $\sigma_j$ to form $\mathbf{X'}$;\label{alg.normalize}
\BState \textit{Get the top} $F$ \textit{singular value matrix} $\mathbf{S}$ \textit{and right-singular vector matrix} $\mathbf{V}$: 
\State $[\mathbf{V},\mathbf{S}]\gets$ svds$(\mathbf{X'})$ via Lanczos algorithm in roughly $O(n_{nz})$ time; \label{alg.svds}
\BState \emph{Adjust maximum eigenvalue}:
\State $\mathbf{m} \gets \mathbf{m}./(\mathbf{\sigma}.\sqrt{n})$;
\State $s_{top}^2 \gets s_{top}^2- \sum_{i=1}^{n}n m_i^2$; $\;\; \lambda_{top}=\sqrt{s_{top}^2}$; \label{alg.shrink}
\BState \emph{Get the cleaned, low-dimensional similarity representation: }
\State $\mathbf{S} \gets \mathbf{V} \times (\mathbf{S}.^2)$;$\;\; \mathbf{V} \gets \mathbf{V}$;\label{alg.lds}
\BState For item $i$ and $j$ the similarity/correlation $c_{ij} = S_i \times V_{j}^T$. 
\EndProcedure
\end{algorithmic}
\end{algorithm}

where, ``$.$'' denotes element-wise operation on vectors and matrices. $S_i$ and $V_j$ denote the $i-th$ and $j-th$ row of the matrices respectively, $s_{top}$ is the largest singular value, $\lambda_{top}$ is the largest eigenvalue and $n_{nz}$ is the number of non-zeros.

Clean-KNN starts by calculating the mean and sum of each column of $\mathbf{X}$ and then it normalizes $\mathbf{X}$ in line \ref{alg.normalize} to form $\mathbf{X'}$. This is so that $\mathbf{X'}^T \mathbf{X'}$ is equal to cosine similarity matrix of $\mathbf{X}$. Since for real matrices the square of the \emph{singular values} of $\mathbf{X'}$ is equal to the \emph{eigenvalues} of $\mathbf{X'}^T \mathbf{X'}$ while the eigenvectors are the same, Clean-KNN just calculates the right-singular vectors and singular values of $\mathbf{X'}$ in line \ref{alg.svds}. In line \ref{alg.shrink} the top eigenvalue is shrunk according to Equation \ref{eqn:overestimate}. Finally, we get the low-dimensional similarity representation in line \ref{alg.lds}. We note that Clean-KNN can also be used for user-user similarity by transposing $\mathbf{X}$.

\section{Experiments}\label{EXP}
We aim to answer the following questions via quantitative evaluation: i) Is noise removed by removing the bulk of the eigenvalues?
ii) Does the shrinkage of $\lambda_{top}$ improve performance? 

For our experiments we used Movielens1M dataset\footnote{\scriptsize \url{https://grouplens.org/datasets/movielens/1m/}}(ML1M) and converted it to implicit feedback \footnote{We focused on implicit feedback since it is closer to the real user behavior and is the focus of most research, however, our results generalize to explicit feedback.}
by ignoring the rating magnitudes. We used four evaluation metrics namely, recall@50 (R@50), normalized discounted cumulative gain (NDCG@50), area under the curve (AUC) and diversity@50 (D@50). D@N is the total number of distinct items in the top-N list across all users. 

\subsection{Baselines and Parameters}
Weighted user-KNN (WUkNN) and weighted item-KNN (WIkNN) were used as the base recommenders, with the similarity function defined by Equation \ref{eq:cosine}. We also compare our performance with a well know item recommender SLIM \cite{ning2011slim}, and the vanilla SVD recommender (svds in MATLAB) which used the same number of factors $F$ as Clean-KNN. We performed 5-fold cross-validation to select the parameters. We searched for $\lambda_{cut}$ by incrementing $F$ by $10$ when $10 \leq F \leq 100$ and in increments of $100$ afterwards till we reach close to $\lambda_{max}$.
\section{Results}\label{Result}
The results are shown in Table \ref{tab:mainResult}. It is worth mentioning here that we do not aim to provide state of the art results, rather we aim to gain insights into the similarity metrics used by memory based methods and demonstrate the effects of these insights on the performance. We note that Clean-KNN improves the performs over the vanilla kNN. We also see that it is better than vanilla SVD with the same number of factors. 
\subsection{Is noise removed?}
For both datasets, the table is divided into subsections by dashed horizontal lines. In each subsection we want to highlight two scenarios: (a) the best base KNN recommender, and (b) the noise removed Clean-KNN recommender of Algorithm \ref{algo}. We can see that the performance of the scenario (b) is better than scenario (a). This signifies that most of the removed eigenvalues did not carry much useful information and hence can be categorized as noise.

\subsection{Does shrinkage of $\lambda_{top}$ help?}
To answer this question we have to compare a base user or item-KNN recommender with a recommender that contains all the eigenvalues but shrinks the top eigenvalue according to Equation \ref{eqn:overestimate}. Note, that this recommender is created for illustration of the effectiveness of the shrinkage procedure. The performance of this recommender is shown in Table \ref{tab:mainResult} and labeled as (c). We see that the performance of the scenario (c) is always better than scenario (a). This confirms that just by shrinking $\lambda_{top}$ we get improved performance. In addition, scenario (c) is still outperformed by scenario (b), thus this confirms the utility of the clipping strategy.

 
\sisetup{detect-weight=true,detect-inline-weight=math} 
\begin{table}[]
\sisetup{round-mode=places}
\centering
\sisetup{round-mode=places}
\caption{Performance of Clean-KNN w.r.t. four metrics shows that it outperforms its vanilla counterparts.}
\label{tab:mainResult}
{\small
\begin{tabular}{lS[round-precision=3]S[round-precision=3]S[round-precision=3]l}
\hline
{\textbf{\scriptsize{Movielens1M}}}                       & {\textbf{\scriptsize{NDCG@50}}} &  {\textbf{\scriptsize{AUC} }}    & {\textbf{\scriptsize{R@50}}} &{\textbf{ \scriptsize{D@50}}} 
\\
\hline
\scriptsize{(a)WUKNN}\tiny{($k=500$)}                               & 0.34513	&0.90509	&0.34577	&661       
 \\
\scriptsize{(b)\textbf{Clean-UKNN}}\tiny{($k=500, F=400$)}                   & 0.36077	&0.91152&	0.36395	&761        
\\
\scriptsize{(c)Shrink-UKNN}\tiny{($k=500$)}                   &0.35826	&0.91105	&0.3608	&720
     
  \\ \hdashline
\scriptsize{(a)WIKNN}\tiny{($k=500$)}                               & 0.35629	&0.9115&	0.35458&	1668
       
  \\ 
\scriptsize{(b)\textbf{Clean-IKNN}}\tiny{($k=500, F=400$)}                   &0.36834	&0.9193	&0.37833	&2187
      
  \\
  \scriptsize{(c)Shrink-IKNN}\tiny{($k=500$)}                   & 0.36938	&0.91664&	0.3678	&1730

  \\
 \hdashline
\scriptsize{SVD}\tiny{($F=400$)}                                & 0.23632&	0.77011	&0.24779	&2242   
  \\

\scriptsize{SLIM}\tiny{($L_1=10^{-2}, L_2=10^{-3},k=500$)} &0.29339	&0.88235&	0.30035	&534
       
\\
\hline

\end{tabular}
}
\end{table}

\section{Conclusion}\label{conclusion}
Memory-based recommenders are one of the earliest recommendation techniques which are still being deployed in the industry today in conjunction with other methods. In this paper, we analyzed the spectral properties of the Pearson and cosine similarities. And we used insights from MP-law to show that these empirical similarities suffer from noise and eigenvalue spreading. We showed that the cosine similarity naturally performs the eigenvalue shrinkage but it overestimates $\lambda_{top}$. We then provided a linear time and memory efficient cleaning strategy, Clean-KNN, that removes noise and corrects for the overestimation of $\lambda_{top}$. Through empirical evaluation, we showed that this cleaning strategy is effective and results in better performance, in terms of accuracy and diversity, compared to the vanilla kNN recommenders.

\bibliographystyle{ACM-Reference-Format}
\bibliography{RMT} 

\end{document}